\documentclass[11pt]{JHEP3}
\newcommand{\Comment}[1]{{}}
\usepackage{amsfonts,amsthm,amsmath,amssymb,slashed}

\newcommand\nn{\nonumber}

\newcommand{\beq}{\begin{equation}}
\newcommand{\eeq}{\end{equation}}

\newcommand{\beqa}{\begin{eqnarray}}
\newcommand{\eeqa}{\end{eqnarray}}

\newcommand{\bn}{\begin{equation}}
\newcommand{\en}{\end{equation}}
\newcommand{\by}{\begin{eqnarray}}
\newcommand{\ey}{\end{eqnarray}}

\newcommand{\de}{\delta}
\newcommand{\ep}{\epsilon}
\newcommand{\om}{\omega}
\newcommand{\si}{\sigma}

\newcommand\fverb{\setbox\fverbbox=\hbox\bgroup\verb}
\newcommand\fverbdo{\egroup\medskip\noindent%
            \fbox{\unhbox\fverbbox}\ }
\newcommand\fverbit{\egroup\item[\fbox{\unhbox\fverbbox}]}
\newbox\fverbbox

\newcommand{\nablaslash}{\not{\hbox{\kern-3pt $\nabla$}}}

\title{Aspects of topologically gauged M2-branes with six supersymmetries: towards  a "sequential AdS/CFT"?
 }

\author{Bengt E.W.~Nilsson\footnote{This note is based on  talks at 35'th Johns Hopkins Workshop, Budapest, June 22-24, 2011, "QTS7", Prague, August 7-13, 2011, "Geometry of strings and fields", Nordita, Stockholm, November 15, 2011 and "Iberian Strings 2012", Bilbao,
January 31-Febraury 2, 2012.}%\\
\\\\
Fundamental Physics\\
Chalmers University of Technology\\
SE-412 96 G\"oteborg, Sweden\\

{\tt {\footnotesize  tfebn@chalmers.se}}}

\abstract{In this note we review the construction of topologically gauged M2 branes with 6 supersymmetries and discuss some of its properties.
This is done using  the 3-algebra formulation thereby covering all   possible gauge groups. 
We will elaborate upon 1)  the fundamental identity and its solutions
noting, provided these gauged theories describe stacks of branes,  the case of a single brane, 
2) the chiral point solution to the field equations (occuring even for a single brane)
 that breaks the superconformal symmetries down to those of
$AdS_3$ (TMG) supergravity,  3)  physical parameters and how
they scale  in the compactification from 11d to 10d and give rise   to matter theories in curved space-times, 
and finally 4) a more speculative comment on "$sequential$ $AdS/CFT$".
Here we propose that the superconformal symmetry breaking in topologically gauged theories leads  to the sequence 
$AdS_4/CFT_3 \rightarrow AdS_3/CFT_2$ and  that the higgsing in the 3d boundary 
theory is related to a change of foliation in the 
$AdS_4$ bulk theory. }
 

\begin{document}

%\maketitle  IS IGNORED %%%%%%%%%%%

\setcounter{page}{2}

\section{Introduction and summary}

At the IR fix-point stacks of M2-branes are  described by interacting superconformal Chern-Simons(CS) matter theories \cite{Schwarz:2004yj}.
Such theories were first discovered with eight supersymmetries by Bagger and Lambert and independently by Gustavsson (BLG)
\cite{Bagger:2006sk,Gustavsson:2007vu,Bagger:2007jr,Bagger:2007vi}
and were shown to have a number of interesting properties. It was, however,  soon realized that this (classical) theory has the problem   that it 
is heavily constrained with only  one possible solution of the fundamental identity, or in other words the theory exists only for one gauge group, namely $SO(4)$.
Although a level $k$ can be introduced (by rescaling the four-index structure constants) and argued to take any integer value, only $k=1,2$ have a possible interpretation in terms of stacks of M2-branes. One is therefore forced in the M-theory context to deal with a strongly coupled system since the weak coupling limit requires taking $k$ large.
The BLG theory seems for this reason to describe only stacks of two branes although more recent work involving monopole operators \cite{Aharony:2008ug}
has produced a number of new quantum theories with eight supersymmetries, see e.g. \cite{Bashkirov:2011pt} and references therein. 

Reducing the number of supersymmetries to six gives rise to the so called ABJM theories \cite{Aharony:2008ug} which already at the classical level are much more general and can describe stacks with any number 
of branes and with a  level that can take any integer value. In this case the interpretation in terms of stacks of M2-branes is clear and so is its
connection to eleven dimensional supergravity and M-theory compactified on $AdS_4\times S^7/{\bf Z}_k$ where $k$ is the level in the ABJM  $CFT$ \cite{Aharony:2008ug} which is based on the  construction in \cite {Nilsson:1984bj}. 
The much looser structure found here makes the generalization
to other gauge groups than those discussed by ABJM possible \cite{Aharony:2008gk}. The admissible gauge groups have been completely classified using 
a variety of methods, 
see \cite{Papadopoulos:2008sk, Gauntlett:2008uf, Schnabl:2008wj, Palmkvist:2009qq}. In particular the work by Palmkvist  \cite{Palmkvist:2009qq} is relevant here since 
the method used there is based on the three-algebra formulation of the $N=6$ theories.

The above superconformal M2-brane theories have apart from the gauge symmetries associated with the Chern-Simons (CS) vector gauge fields
also global symmetries corresponding to the superconformal group namely translations, supersymmetry and their $special$ conformal counterparts plus Lorentz rotations, dilatation and R-symmetries. We will here describe how these global symmetries can be made local by introducing conformal supergravity into the game and 
couple it to the BLG \cite{Gran:2008qx} and ABJM  \cite{Chu:2009gi} CS matter theories. While the construction in the case of BLG met with difficulties and could not be carried through completely in \cite{Gran:2008qx}\footnote{The complete theory has now been constructed \cite{Gran:2012}.}, the 
complete lagrangian for the ABJM case has, however, been derived in  \cite{Chu:2009gi}. As we will see below it is found to have new potential terms for the scalar fields over and above those present in the ungauged ABJM theory. This leads to some very interesting new properties in particular one finds  a Higgs effect with an 
intriguing end result  related to chiral gravity  \cite{Chu:2009gi}. 

After a brief discussion of the topologically gauged theories and how they are constructed, we turn to some of their special properties related to the new potential terms  that  were found in  the ABJM topologically gauged theory in  \cite{Chu:2009gi} and worked out in more detail in some special cases in \cite{Chu:2010fk}. We will see that the new structure of the scalar potential leads to a Higgs effect that breaks the 
superconformal symmetries to those of $AdS_3$ corresponding to a compactification from M-theory in eleven dimensions to string theory in ten. The end-result is then a non-trivial interacting CS matter theory in a curved background. We will also 
 try to argue that theories of this type may arise from $AdS_4/CFT_3$ based on Neumann boundary conditions as has been advocated, e.g., in work by Marolf, de Haro and others  \cite{Marolf:2006nd, Compere:2008us, Amsel:2009rr, deHaro:2008gp}. Since the presence of new terms in the scalar field potential leads to a Higgs effect which breaks the conformal symmetries to those of $AdS_3$ it  suggests the possibility that a second $AdS/CFT$ comes into play relating the theory in $AdS_3$ to a $CFT_2$, something that might be called "sequential $AdS/CFT$":
\beq
AdS_4(N)/CFT_3(TG)\rightarrow AdS_3(H)/CFT_2
\eeq
where $N$ refers to Neumann boundary conditions, $TG$ to topologically gauged BLG or ABJM theory and $H$ to its higgsed version. The final $CFT_2$ is unknown. As we will see in the discussion of the structure constants later this scenario is also possible for the case of a single brane.
The assumptions needed for this sequential $AdS/CFT$ to work will be discussed in more detail in a later section.

 A similar idea was put forward some time ago by M. Vasiliev \cite{Vasiliev:2001zy}\footnote{I am grateful to M. Vasiliev for discussions about these ideas.} based on properties of higher spin (HS) algebras but  without having
any theory that explicitly can generate the necessary breaking from conformal to AdS symmetries as  we have in the present work.
It might be argued that going to the boundary of $AdS$ twice is impossible  because the boundary of a boundary is normally zero. A possible way out of this problem might be to relate the higgsing described above to a change of foliation of the $AdS_4$  space. Indeed it is well-known that $AdS$ spaces can be foliated in many ways some of them leading to leaves that are themselves $AdS$ spaces, see e.g.  \cite{Emparan:1999pm, Jatkar:2011ue}. A recent discussion of phenomena of this kind can be found in 
 \cite{Andrade:2011nh} where also the possibility of a sequential $AdS/CFT$ is mentioned. This is  based on previous work in \cite{Compere:2008us}.

\section{Construction of topologically gauged M2 brane theories with six supersymmetries}

The title of this section refers to the theories  obtained by gauging all global symmetries of the ABJM/ABJ type theories \cite{Aharony:2008ug, Aharony:2008gk} keeping the superconformal symmetries
intact but  turned into local ones.
This can be achieved by coupling, for instance using a Noether procedure, the Chern-Simons-like theory for $N=6$ conformal supergravity to the ABJM/ABJ
 CS matter theories as first done in the BLG case in \cite{Gran:2008qx}. The only non-trivial aspect of this construction is that, in order to get a supersymmetric lagrangian,  one has to add a $U(1)$ gauge field not a priori present in either the on-shell superconformal supergravity lagrangian or in the  ABJM/ABJ matter theories.  The details of the whole construction can be found in \cite{Chu:2009gi}\footnote{The topologically gauged lagrangian below has so far been derived only in this way. However,  there are still some multifermionic terms in the variation of the lagrangian that have not been verified to cancel. Although it is very unlikely that something could go wrong at the final steps of the construction it would be welcome to have an alternative derivation in order to prove that the theory exists. Work in this direction is currently under way  \cite{Gran:2012}}. It should be noted that the  topological gauging leads to the an interacting theory even if we start from a free theory of scalars and spinors. In particular the $CFT$ for a single brane will start to self-interact when it is gauged.
 
 The on-shell supergravity fields are
 the
three gauge fields of 'spin' 2, 3/2 and 1, i.e.
\beq
e_{\mu}{}^{\alpha},\,\,\chi_{\mu AB},\,\,B_{\mu}^{A}{}_B
\eeq
The capital Latin indices used here are in the fundamental of the R-symmetry group $SU(4)\times U(1)$
and the corresponding gauge field $B_{\mu}^{A}{}_B$ is  in the adjoint
of $SU(4)$ while the $ U(1)$ factor plays no role in the gravity sector. It might be argued that the extra gauge field
$C_{\mu}$ that one is forced to introduce in the coupling to matter is just the gauge field of this $ U(1)$ factor. It is in fact part of the off-shell multiplet \cite{Howe:1995zm},
see also \cite{Cederwall:2011pu}.
The spin 3/2 field $\chi_{\mu AB}$, and the supersymmetry parameter $\ep_{AB}$, are both antisymmetric and self-dual in  the two indices $AB$
in order to accommodate exactly six supersymmetries. The matter sector contains the fields
\beq
Z^A_a,\,\,\,\Psi_{Aa},\,\,\,\tilde A_{\mu}{}^a{}_b=A_{\mu}{}^d{}_c f^{ac}{}_{bd}
\eeq
where one should remember that the complex conjugation acts on all the indices\footnote{The space-time  spinors 
are not affected by complex conjugations since they are Majorana.} as $(Z^A_a)^*=\bar Z^a_A$ and $(\Psi_{Aa})^*=\Psi^{Aa}$, see \cite{Nilsson:2008kq}.
The three-algebra formulation was originally obtained from the ABJM quiver version in \cite{Bagger:2008se}.
While the BLG and ABJM Chern-Simons theories are well studied in a large number of papers 
that have appeared since their introduction in 2007, the conformal supergravity theories that we need here are perhaps 
less familiar. These were originally constructed in the 1980's  by Deser and Kay \cite{Deser:1982sw}, van
Nieuwenhuizen \cite{VanNieuwenhuizen:1985ff} for $N=1$, and by Lindstr\"om and
Ro\v cek \cite{Lindstrom:1989eg} for arbitrary $N$. 

Using the Noether procedure we find that the coupled lagrangian of the topologically gauged theory is given by (with $A^2=\tfrac{1}{2}$) \cite{Chu:2009gi, Chu:2010fk}
\begin{eqnarray}
 L&=&g_M^{-2}L_{sugra}^{conf}+L_{ABJM}^{cov}+\tfrac{1}{2g_M^2}\epsilon^{\mu\nu\rho}C_{\mu}
 \partial_{\nu}C_{\rho}\label{CS:no idex}\cr
  &&+iAe\bar\chi_{\mu}^{BA}\gamma^{\nu}\gamma^{\mu}\Psi_{Aa}
       (\tilde D_{\nu}\bar Z^a_B-\tfrac{i}{2} A\bar\chi_{\nu BC}\Psi^{Ca})+c.c.\label{squared:1}\cr
  &&+i \epsilon ^{\mu\nu\rho} (\bar\chi_{\mu}^{AC}\chi_{\nu BC}) Z^B_a
       \tilde D_{\rho} \bar Z^a_A+c.c.\label{squared:2}\cr
 &&-iA(\bar f^{\mu AB}\gamma_{\mu}\Psi_{Aa}\bar Z_B^a+
 \bar f^{\mu}_{AB}\gamma_{\mu}\Psi^{Aa}Z^B_a)\label{squared:3}\cr
 &&-\tfrac{e}{8}\tilde R \vert Z\vert^2+\frac{i}{2}\vert Z \vert^2\bar f^{\mu}_{AB}\chi_{\mu}^{AB}\label{squared:4}\cr
 % &&-ieA f^{ab}{}_{cd}(\bar \chi_{\mu AB}\gamma^{\mu}\Psi^{Dd})Z^A_a Z^B_b \bar Z^c_D+c.c.\notag
 % (the last  line is one of the terms in the previous line when breaking up the antisymmetry!  )\\
 &&+2i\lambda A e f^{ab}{}_{cd}(\bar \chi_{\mu AB}\gamma^{\mu}\Psi^{d[B})Z^{D]}_a Z^A_b \bar Z^c_D+c.c.\label{hatted:B}\cr
 &&-i\lambda\epsilon^{\mu\nu\rho}(\bar \chi_{\mu AB}\gamma_{\nu}\chi_{\rho}^{CD})(Z^A_a Z^B_b\bar Z^c_C \bar Z^d_D)f^{ab}{}_{cd}\notag\cr
 &&+\frac{i\lambda}{4}\epsilon^{\mu\nu\rho}(\bar\chi_{\mu AB}\gamma_{\nu}\chi_{\rho}^{AB})
 (Z^C_a Z^D_b \bar Z^c_C \bar Z^d_D)f^{ab}{}_{cd}\label{hatted:C}\cr
 &&-\frac{ig_M^2}{16}e \epsilon^{ABCD}(\bar\Psi_{Aa}\Psi_{Bb})\bar Z^a_C\bar Z^b_D+c.c.\notag\cr
 &&+\frac{ig_M^2}{16} e(\bar\Psi_{Db}\Psi^{Db})\vert Z\vert^2-\frac{ig_M^2}{4}e(\bar\Psi_{Db}\Psi^{Bb})\bar Z^a_B Z^D_a\notag\cr
 &&+\frac{ig_M^2}{8}e(\bar \Psi_{Db}\Psi^{Da})\bar Z^b_B Z^B_a
   +\frac{3ig_M^2}{8}e (\bar \Psi_{Db}\Psi^{Ba})\bar Z^b_B Z^D_a \label{hatted:D}\cr
 &&-\frac{ig_M^2}{16}eA(\bar \chi_{\mu AB} \gamma^{\mu}\Psi^{Bb}) \vert Z\vert^2 Z_b^A
 -\frac{ig_M^2}{4}eA(\bar \chi_{\mu AB} \gamma^{\mu}\Psi^{Db}) Z^A_a Z^B_b \bar Z^a_D+c.c\label{hatted:E}\cr
 &&-\frac{ig_M^2}{4}\epsilon^{\mu\nu\rho}(\bar \chi_{\nu AB} \gamma_{\rho}\chi_{\mu}^{CD}) Z^A_a Z^B_b \bar Z^a_C \bar Z^b_D
  +\frac{ig_M^2}{64}\epsilon^{\mu\nu\rho}(\bar \chi_{\nu AB} \gamma_{\rho}\chi_{\mu}^{AB})\vert Z\vert^4 \label{hatted:F}\cr
 &&+\frac{\lambda g_M^2}{8}ef^{ab}{}_{cd} \vert Z\vert^2 Z^C_a Z^D_b \bar Z^c_C \bar
          Z^d_D+\frac{\lambda g_M^2}{2}e f^{ab}{}_{cd}Z^B_a Z^C_b Z^D_e \bar Z^e_B \bar Z^c_C \bar Z^d_D\label{V:prm}\cr
 &&+\frac{5g_M^4}{12 \cdot 64}e(\vert Z\vert^2)^3 -\frac{g_M^4}{32}e \vert Z\vert^2 Z^A_b Z^C_a \bar Z^b_C \bar Z^a_A+
 \frac{g_M^4}{48}e Z^A_a Z^B_b Z^C_d \bar
          Z^b_A \bar Z^d_B \bar Z^a_C,\label{V:prmprm}
  \end{eqnarray}
   where   $|Z|^2$ stands for
\beq
Z^A_a\bar Z_A^a=Tr(\bar Z_A,Z^A)\;,
\eeq
  and $c.c.$ refers to complex conjugation of the term on the line where it occurs.
  
  For the two subsectors that are coupled we use the following lagrangians 
\beqa
L^{cov}_{ABJM}&=& -e(\tilde D_\mu Z^{A}{}_a)(\tilde D^\mu \bar{Z}_A{}^a) - \frac{1}{2}(ie \bar \Psi^{Aa} \gamma^\mu  \tilde D_\mu  \Psi_{Aa}+
ie \bar \Psi_{Aa} \gamma^\mu \tilde D_\mu  \Psi^{Aa})
 \cr
 &&  - ie\lambda f^{ab}{}_{cd}\bar {\Psi}^{Ad}
 \Psi_{Aa}Z^B{}_b \bar{Z}_{B}{}^{c}+
 2ie\lambda f^{ab}{}_{cd}\bar {\Psi}^{Ad}  \Psi_{Ba}Z^B{}_b \bar{Z}_{A}{}^{c}
  \cr
 &&
 -\tfrac{i}{2}\lambda e \epsilon_{ABCD} f^{ab}{}_{cd}\bar {\Psi}^{Ac}  \Psi^{Bd} Z^C{}_a Z^D{}_b
  -\tfrac{i}{2}\lambda e \epsilon^{ABCD} f^{cd}{}_{ab}\bar {\Psi}_{Ac}  \Psi_{Bd}\bar{Z}_{C}{}^a \bar{Z}_{D}{}^{b}
\cr
 && -eV +\tfrac{1}{2}\lambda\epsilon^{\mu\nu\lambda}(
f^{ab}{}_{cd}A_{\mu}{}^d{}_{b}
\partial_\nu A_{\lambda}{}^c{}_{a}+ \tfrac{2}{3}\lambda f^{bd}{}_{gc} f^{gf}{}_{ae}
A_{\mu}{}^a{}_{b}  A_{\nu}{}^c{}_{d} A_{\lambda}{}^e{}_{f})\,,
\\
V &=& \tfrac{2}{3} \Upsilon^{CD}{}_{Bd}\bar\Upsilon_{CD}{}^{Bd}\,,\\
\Upsilon^{CD}{}_{Bd}&=&\lambda f^{ab}{}_{cd} Z^C{}_a{Z}^D{}_b \bar{Z}_B{}^c
  +\lambda f^{ab}{}_{cd}\delta^{[C}{}_B Z^{D]}{}_a{Z}^E{}_b \bar{Z}_{E}{}^{c}
\eeqa
and
\beqa
L^{conf}_{sugra} &=&\frac{1}{2}\varepsilon^{\mu\nu\rho}
Tr_{\alpha}(\tilde\omega_{\mu}\partial_{\nu}\tilde\omega_{\rho}+
\frac{2}{3}\tilde\omega_{\mu}\tilde\omega_{\nu}\tilde\omega_{\rho})
-2\varepsilon^{\mu\nu\rho}Tr_A
(B_{\mu}\partial_{\nu}B_{\rho}+\frac{2}{3}B_{\mu}B_{\nu}B_{\rho})\\\nn
&&-i e^{-1} \varepsilon^{\alpha\mu\nu}\epsilon^{\beta\rho\sigma}(\tilde
D_{\mu}\bar{\chi}^{AB}_{\nu}\gamma_{\beta}\gamma_{\alpha}\tilde
D_{\rho}\chi_{\sigma AB}).
\eeqa
and the action of the covariant derivative on a spinor is given by
\beq
\tilde D_{\mu}\Psi^{Aa}=\partial_{\mu}\Psi^{Aa}+\tfrac{1}{4}\tilde
\omega_{\mu \alpha\beta}\gamma^{\alpha\beta}\Psi^{Aa}+B_{\mu
B}^A\Psi^{Ba}+\lambda\tilde A_{\mu
b}^a\Psi^{Ab}+qC_{\mu} \Psi^{Aa}.\label{Ccovderivative}
\eeq
From the work of \cite{Chu:2009gi}\footnote{In this reference this result is presented as  $q^2=\tfrac{1}{16}$ but it is more appropriate to use $"q"$ instead of its square as done in this presentation.}  we know that the $U(1)$ charge $q=\tfrac{1}{16}$ in order for the theory to possess six local special conformal as well as  ordinary supersymmetries. Furthermore, the level $k$ or its inverse $\lambda=\tfrac{2\pi}{k}$ can be introduced as usual by rescaling the structure constant but besides that
one can also, as done in \cite{Chu:2010fk}, introduce a 
 dimensionless gravitational coupling constant  $g_M$ by rescaling the trace in the three-algebra. Another, perhaps more physical way, to get  $g_M$ in the right places in the lagrangian is to, in the Noether construction of \cite{Chu:2009gi},  start from a pure supergravity lagrangian multiplied by a factor $g^2_M$. This will automatically give the correct result.

The supersymmetry variations under which the above lagrangian is invariant are given by
\beqa
 \delta e_{\mu}{}^{\alpha}&=& i \bar\epsilon_{gAB}\gamma^{\alpha}\chi_{\mu}^{AB},\\
 \delta \chi_{\mu}^{AB}&=&\tilde D_{\mu}\epsilon_{g}^{AB},\\
 \delta B_{\mu ~B}^{~A}&=&\frac{i}{e}(\bar f^{\nu AC}\gamma_{\mu}\gamma_{\nu}\epsilon_{g BC}-
  \bar f^{\nu}_{BC}\gamma_{\mu}\gamma_{\nu}\epsilon_{g}^{AC}
             )\cr
   &&+\tfrac{i}{4}g^2_M(\bar\epsilon_{BD}\gamma_{\mu}\Psi^{a(D} Z^{A)}_a-\bar\epsilon^{AD}\gamma_{\mu}\Psi_{a(D} \bar Z_{B)}^a)
         \cr
   &&-\tfrac{i}{2}g^2_M(\bar\epsilon^{AC}_g\chi_{\mu DC}Z^D_a\bar Z^a_B-\bar\epsilon_{g BC}\chi^{DC}_{\mu}Z^A_a\bar Z^a_D)\cr
      && +\tfrac{i}{8}g^2_M\delta^A_B(\bar\epsilon^{EC}_g\chi_{\mu DC}-
       \bar\epsilon_{gDC}\chi^{EC}_{\mu})Z^D_a\bar Z^a_E
       \cr
   &&+\tfrac{i}{8}g^2_M(\bar\epsilon^{AD}_g\chi_{\mu BD}-\bar\epsilon_{gBD}\chi^{AD}_{\mu})
   |Z| ^2,\\
 \delta Z^A_a &=&i\bar\epsilon^{AB}\Psi_{Ba}, \\
 \delta\Psi_{Bd}&=&\gamma^\mu\epsilon_{AB}(\tilde D_\mu Z^A_d
 -iA\bar\chi_{\mu}^{AD}\Psi_{Dd})\cr
   &&+\lambda f^{ab}{}_{cd} Z^C{}_a Z^D_b \bar Z_B^c \epsilon_{CD}-\lambda 
    f^{ab}{}_{cd} Z^A_a Z^C_b \bar Z_C^c\epsilon_{AB}\cr
   &&+\frac{1}{4}g^2_MZ^C_c Z^D_d \bar Z_B^c \epsilon_{CD}+\frac{1}{16} g^2_M  |Z| ^2
   Z^A_d\epsilon_{AB},\\
 \delta \tilde A_{\mu~d}^{~c}&=&-i\lambda(\bar\epsilon_{AB}\gamma_{\mu}\Psi^{Aa}Z^B_b-
 \bar\epsilon^{AB}\gamma_{\mu}\Psi_{Ab}\bar Z_B^a)f^{bc}{}_{ad}\cr
    &&-2i\lambda(\bar\epsilon_g^{AD}\chi_{\mu BD}-\bar\epsilon_{gBD}\chi_{\mu}^{AD})Z^B_b\bar Z^a_A
    f^{bc}{}_{ad},\\
\delta C_\mu &=&
-iqg^2_M(\bar\epsilon_{AB}\gamma_{\mu}\Psi^{Aa}Z^B_a-\bar\epsilon^{AB} \gamma_\mu\Psi_{Aa}\bar
Z_B^a)\cr
&&-2iqg^2_M(\bar\epsilon_g^{AD}\chi_{\mu BD}-\bar\epsilon_{gBD}\chi_{\mu}^{AD})Z^B_a\bar Z^a_A\;,
\eeqa
where $ \epsilon_m^{AB}=A\epsilon_g^{AB}=\epsilon^{AB}$  (with $A^2=\tfrac{1}{2}$). 
  
  In presenting the result of gauging the global symmetries of the M2-brane theories with six supersymmetries  we have chosen to use the
  three-algebra formulation. The reason for this choice is not that it is more fundamental than the ABJM/ABJ quiver formulation (which probably is not the case)
  but rather one of convenience. The structure constants and its fundamental identity encode in a single form all possible gauge groups for $N=6$ as classified by
  \cite{Schnabl:2008wj}. Indeed, the fact that the fundamental identity has exactly the same solutions as the quiver formulation has been shown by 
  Palmkvist in \cite{Palmkvist:2009qq}.
  Some implications of this fact will be discussed further in the next section.
  
  \section{Comments}
  
  In this  section we will elaborate on some of the properties of the topologically gauged M2-brane theories with six supersymmetries
 \cite{Chu:2009gi,Chu:2010fk}  that were presented in some detail in the previous section.
  
   \subsection{The role of the structure constants in the three-algebra formulation}
   
   The structure constants used in  \cite{Bagger:2008se, Nilsson:2008kq} are antisymmetric in both the upper and lower pair of indices and the corresponding fundamental identity reads  
   \beq
   f^{e[a}{}_{dc}f^{b]d}{}_{gh}=f^{ab}{}_{d[g}f^{ed}{}_{h]c}.
   \eeq
   There is a connection between this three-algebra and generalized Jordan triple systems as explained in \cite{Nilsson:2008kq, Palmkvist:2009qq} but this will not be used 
   here\footnote{It is interesting to note that the restriction of the generalized Jordan algebra (GJA) to the three-algebra structures found in M2-brane theories is not coming from the Chern-Simons term. In fact, the Chern-Simons term implies  the GJA identity without restrictions.}. 
   Instead we will
   discuss various forms for the structure constants that  solve the fundamental identity and see what the implications are for the theory\footnote{I am grateful to Jakob Palmkvist for discussions concerning the various forms of the structure constants mentioned here.}. Recall first that
   a general way to obtain structure constants that satisfy the fundamental identity is to first replace each three-algebra index by a pair of indices by using the elements of the three-algebra $T^a$ and $T_a$. Thus the scalar fields become  (suppressing the R-symmetry index):
   \beq
   Z_a\rightarrow Z_a(T^a)_i^{i'}=Z_i^{i'},\,\,\, \bar Z^a\rightarrow\bar Z^a(T_a)^i_{i'}= \bar Z^i_{i'}\,,
   \eeq 
   and then set\footnote{The following form follows directly from the results in \cite{Bagger:2008se}.}
    \beq
   f^{ab}{}_{cd} \rightarrow f^i_{i'}{}^j_{j'}{}_k^{k'}{}_l^{l'}=\de^i_k\de^j_l\de_{j'}^{k'}\de^{l'}_{i'}-\de^i_l\de^j_k\de_{i'}^{k'}\de^{l'}_{j'},
   \eeq
   which is often used to rewrite the triple product  as (using $X_a\rightarrow X_i^{i'}$ etc)
   \beq
   f^{ab}{}_{cd}X_aY_b\bar Z^c\rightarrow [X,Y;\bar Z]_l^{l'}:=(X\bar Z Y)_l^{l'}-(Y\bar Z X)_l^{l'}.
   \eeq
   Note that  the ranges of the two types of indices (primed and unprimed) are not related here and the same is true for the kinds of symmetry groups they transform under.
   This form of the structure constants thus corresponds to quivers with gauge groups like $SU(M)\times SU(N)\times U(1)$
   which  explains how both symmetric quivers like the ABJM theories  and  non-symmetric gauge group pairs as in the ABJ cases can be accommodated.
   One can also eliminate, e.g., the primed indices by letting them take only one value giving
   \beq
     f^{ab}{}_{cd} \rightarrow f^{ij}{}_{kl}=\de^{ij}_{kl}=\de^i_k\de^j_l-\de^i_l\de^j_k,
   \eeq
   which translates into
   \beq
    f^{ab}{}_{cd} X_aY_b\bar Z^c\rightarrow (X\cdot \bar Z)Y_l- (Y\cdot \bar Z)X_l.
   \eeq
    In this case the solution to the fundamental identity corresponds to only one non-abelian simple gauge group factor (times a $U(1)$ factor), or  in other words a (complex) vector model in the language of sigma models. This is interesting since vector models have been studied a lot and, e.g., the fix-point structure is much better understood than for matrix-like quiver theories. This will be discussed  in the last subsection below.
    
    Of course, also the cases in the classification that involve symplectic groups can be accommodated. This is done by choosing the structure constants appropriately, namely
    \beq
    f^{ijkl} = J^{ik} J^{jl} - J^{jk} J^{il} - J^{ij} J^{kl},
    \eeq
    where $J^{ij}$ is antisymmetric.
   
   Finally, a free version of ABJM/ABJ corresponding to one M2-brane with six supersymmetries is obtained if the indices are chosen to have just one component which means that the structure constants  vanish. Of course, the center of mass theory obtained in this case  has additional supersymmetries adding up to eight.    
     
The higgsing to be described below is independent of the structure constants and thus takes place in all the different cases discussed above in particular for the
   theory  describing a single M2-brane. It would be interesting to find out what the proper string/M theory interpretation is of such a higgsing to $AdS_3$ which 
   happens to be  a  chiral point super-TMG theory  as we will see below.

      \subsection{The $AdS_3$ solution and the higgsing to a TMG theory at the chiral point with six supersymmetries}
   
   The bosonic part of the lagrangian reads
   \beq
   L=\tfrac{1}{g^2}L_{CS(\om)}-eg^{\mu\nu}\partial_{\mu}Z^A_a\partial_{\nu}\bar Z^a_A-\tfrac{e}{8} |Z|^2 R-e V_{pot}(Z,\bar Z),
   \eeq
    where $V_{pot}(Z,\bar Z)$ is the six-order scalar potential and $L_{CS(\om)}$ the gravitational Chern-Simons term expressed in terms of the spin-connection $\om(e)$.
   By varying the action with respect to the complex scalar fields $\bar Z^a_A$ we get the Klein-Gordon equation
   \beq
   \Box Z^A_a-\tfrac{1}{8}R Z^A_a-\partial_{\bar Z^a_A} V_{pot}(Z,\bar Z)=0,
   \eeq
   while a variation with respect to the dreibein (or the metric) gives the Cotton equation
   \beqa
  &&  \tfrac{1}{g^2}C_{\mu\nu}- \tfrac{e}{8}(R_{\mu\nu}-\tfrac{1}{2}g_{\mu\nu} R)|Z|^2+
 \tfrac{e}{2}g_{\mu\nu}V_{pot}\cr
  && -e(\partial_{\mu}Z^A_a\partial_{\nu}\bar Z^a_A-\tfrac{1}{2}g_{\mu\nu}g^{\rho\sigma}\partial_{\rho}Z^A_a\partial_{\si}\bar Z^a_A)+
   \tfrac{e}{8}(D_{(\mu}D_{\nu)}|Z|^2-g_{\mu\nu}\Box |Z|^2)=0.
  \eeqa
  If we trace the Cotton equation we can use the fact that the Cotton tensor has zero trace to get
  \beq
  \tfrac{1}{8} |Z|^2 R + 3 V_{pot}=\tfrac{1}{2}( Z^A_a\Box \bar Z^a_A+ \bar Z^a_A\Box Z^A_a).
  \eeq
  Next we use the Klein-Gordon equation above to replace the RHS of  the last equation by terms involving the curvature scalar and the potential. We find
    \beq
3 V_{pot}=\tfrac{1}{2} (Z\partial_Z+\bar Z\partial_{\bar Z})V_{pot}.
  \eeq
 Since the potential is homogeneous of degree three in both $Z$ and $\bar Z$
  this equation becomes an identity showing that the KG and Cotton equations are compatible.
  
 We can then easily verify that the above bosonic equations are solved by a scalar VEV  $v$, i.e.,  $<Z^A_a>=v\de^{A4}\de_{a8}$, and an $AdS_3$  background space-time satisfying
 \beq
 R_{AdS}=-24v^{-2}V_{pot}(v).
 \eeq
The extended potential appearing in the topologically gauged theory
is crucial for the rest of the discussion so we present it in some detail here.
Recall that there are terms of this kind already in the ungauged ABJM theory
 \begin{equation}
V^{(st)}_{ABJM}=\tfrac{2e}{3}|\Upsilon^{CD}{}_{Bd}|^2,\,\,
\Upsilon^{CD}{}_{Bd}=\lambda f^{ab}{}_{cd}Z_a^CZ_b^D\bar Z^c_B+\lambda f^{ab}{}_{cd}\delta^{[C}_BZ^{D]}_aZ^E_b\bar Z^c_E \,.\nonumber
\end{equation}
In the three-algebra formulation these terms have two structure constants and correspond to a "single trace"  \cite{Chu:2010fk} in the three-algebra. The new terms arising in the topological gauging  \cite{Chu:2009gi} have either  one structure constant corresponding to a "double trace" (dt)
\begin{equation}
V^{(dt)}_{ABJM}=\tfrac{e}{8}\lambda g^2_M f^{ab}{}_{cd}|Z|^2Z^C_aZ^D_b\bar Z^c_CZ^d_D+\tfrac{e}{2}\lambda g^2_M f^{ab}{}_{cd}Z^B_aZ^C_b(Z^D_e\bar Z_B^e)\bar Z_C^c\bar Z_D^d \,.
\end{equation}
or   no structure constant corresponding to a "triple trace" (tt)
\begin{equation}
V^{(tt)}_{ABJM}=\tfrac{5e}{12\times 64}g^4_M(|Z|^2)^3-\tfrac{e}{32}|Z|^2|Z|^4+\tfrac{e}{48}|Z|^6\,,
\end{equation}
where the double trace term $|Z|^4=(Z_a^A\bar Z^a_B)(Z_b^B\bar Z^b_A)$ etc.  One should note how the two parameters $\lambda$ and $g_M$ appear in these expressions  \cite{Chu:2010fk}. 

The derivation of the lagrangian in this case has been checked for all terms in $\delta L$ except a few multi-fermion non-derivative terms. The calculation involves a large number of cross checks on the coefficients appearing in both the lagrangian and 
extended transformation rules. Besides this fact the properties possessed by the last set of six-order potential terms for the complex ABJM scalar fields above
strongly indicate that the construction is correct. In fact, collecting the relevant terms and inserting the VEV for the scalar fields  \cite{Chu:2010fk}
\beq
Z^A=v \delta^A_4+z^A,
\eeq
where for simplicity we have given this equation in the specific case of the $U(N)\times U(N)$ quiver version \cite{Chu:2010fk}, that is, the VEV term is also diagonal in the two fundamental indices leading to an identification of the two gauge groups as in the original version of this Higgs effect \cite{Mukhi:2008ux}.
 
 We can now evaluate the potential for this VEV, where only the terms in the potential not containing structure constants contribute. We then find that
 \beq
  R_{AdS}=-24v^{-2}V_{pot}(v)=-24v^{-2}(\tfrac{1}{16^2}g_M^4v^6)=-\tfrac{3}{32}(g_Mv)^4,
 \eeq
 which means that the three-dimensional cosmological constant is
 \beq
  \Lambda=-\tfrac{1}{64}(g_Mv)^4.
  \eeq

  Inserting the scalar VEV into the above  bosonic lagrangian we find
    \beq
   L=\tfrac{1}{g_M^2}L_{CS(\om)}-\tfrac{e}{8} v^2 R-e (\tfrac{1}{16^2}g_M^4v^6).
   \eeq
   By  introducing the Newton's constant  through $\tfrac{1}{8} v^2=\kappa^{-2}$ and the cosmological length scale $l$ by $\Lambda=-\tfrac{1}{l^2}$ we can  compare our situation  to that of  Li, Song and Strominger \cite{Li:2008dq} in their analysis of chiral gravity in three dimensions.  However, since we have  opposite signs for the Einstein-Hilbert and cosmological terms we should use the following lagrangian instead
     \beq
   L=\tfrac{1}{\kappa^2}(\tfrac{1}{\mu}L_{CS(\om)} -(  R-2\Lambda)).
   \eeq
   If we read of the values of these parameters in terms of our parameters $v$ and $g_M$ we find
   \beq
   \mu=\tfrac{g_M^2}{\kappa^2}=\tfrac{1}{8} (g_Mv)^2,\,\,\,l^{-2}=\tfrac{1}{64}(g_Mv)^4,
   \eeq
     i.e., 
     \beq
     \mu l= 1
     \eeq
     and the theory has thus been higgsed \cite{Chu:2009gi, Chu:2010fk} into a topologically massive AdS supergravity  theory  (TMG)  \cite{Deser:1981wh} at the chiral point.
 This means that one needs to understand the implications of the negative energy black holes that  appear as solution in this kind of theories  and refer the reader to a recent discussion of some related issues by Deser and Franklin \cite{Deser:2010df}.
     
     In the higgsed $AdS$ phase  there are six residual supersymmetries.  These are obtained after the breaking of the superconformal symmetries as linear combinations
     of the original supersymmetries and superconformal symmetries. The combination chosen properly can be seen to generate the correct $AdS$ covariant derivatives and to eliminate a term that otherwise would have been a  Goldstone term in the transformation rules.
     
     One should be able to  follow all the degrees of freedom in the  $AdS_3$ phase of the theory backwards to the pre-higgsed superconformal phase to conclude that 
     there can not be any propagating gravitational modes in $AdS_3$  after the higgsing. This might, in fact,  be considered to be the reason why the theory ends up at 
     chiral point. At the chiral point there are problematic log-modes but off  the chiral point there would have appeared massive gravitational modes that seem 
     difficult to account for in our present situation. We have nothing to add concerning the log-modes but it might interesting to study this issue in light of the connection of the chiral point
     to a conformal theory that one finds in the topologically gauged $N=6$ M2-brane theories studied here. If this argument is correct one would expect the same chiral point phenomenon to occur in the topologically gauged BLG theory  partly constructed in \cite{Gran:2008qx}. The answer to this question has, however, to await the construction of the complete lagrangian in that case (see  \cite{Gran:2012}) and a proper analysis of the nature of the degrees of freedom. 
     
     In somewhat more detail the BLG situation is currently as follows.
     The coupling of the superconformal gravity theory to BLG was attempted in \cite{Gran:2008qx} using the Noether method. The  
calculations can preferably be organized in number of derivatives appearing in the terms  in the supervariation of the lagrangian.
The virtue of this approach is that the variations of the Chern-Simons vector gauge fields need not be assumed but can 
instead be directly inferred from the calculation where the only input is the expressions for the variation of the matter fields (i.e spin 0 and 1/2). One then finds that it is necessary to introduce a number of new terms in the coupled lagrangian beyond the
standard Noether  term  coupling  the supercurrent and the Rarita-Schwinger field.
For BLG this can be carried through in \cite{Gran:2008qx} cancelling all terms in $\delta L$ at order three and two in covariant derivatives. At this order in derivatives
the result was
\begin{eqnarray}
L_{BLG}^{top}&=& L^{conf}_{grav}+L^{cov}_{BLG}\cr
&+& \tfrac{1}{\sqrt 2}ie \bar\chi_{\mu}\Gamma^i
\gamma^{\nu}\gamma^{\mu}\Psi^a \tilde
 D_{\nu}X^{ia}\cr
 %-\tfrac{i}{2\sqrt2}\bar\chi_{\nu}\Gamma^i \Psi^a)
%+\tfrac{1}{6\sqrt2}ie\bar\chi_{\mu}\gamma^{\mu}\Gamma^{ijk}\Psi_a(X^i_bX^j_cX^k_d)f^{abcd}\cr
% \tfrac{1}{48\sqrt2}e\bar\chi_{\mu}\Gamma^{ijkl}\chi^{\mu}(X^i_aX^j_bX^k_cX_d^l)f^{abcd}\nonumber
&-&\tfrac{i}{4}\epsilon^{\mu\nu\rho}\bar\chi_{\mu}\Gamma^{ij}\chi_{\nu}(X_a^i \tilde D_{\rho}X_a^j)+ \tfrac{i}{\sqrt2}\bar
f^{\mu}\Gamma^i\gamma_{\mu}\Psi_a X_a^i\cr
&-&\tfrac{e}{16}X^2\tilde R +\tfrac{i}{16}X^2 \bar
f^{\mu}\chi_{\mu},
\end{eqnarray}
where $i$ is an 8-dimensional vector R-symmetry index and $f^{abcd}$ is the totally antisymmetric three-algebra structure constants.
$f^{\mu}$ is the dualized Rarita-Schwinger field strength of the "spin 3/2" field $\chi_{\mu}$.  

Interestingly enough the last two terms are exactly as expected from ordinary local scale invariance giving some 
hopes that this construction actually makes sense. It would be most welcome to find another approach to derive  this theory.
In \cite{Cederwall:2011pu} superspace methods\footnote{This is closely related to previous work in \cite{Howe:1995zm}. See also \cite{Kuzenko:2011xg, Greitz:2011vh} for some more recent work relevant in this context.} were adopted but this has so far not led to any conclusive results concerning the existence of this theory.
Since as we will see below,  the ABJM construction works without any problems, one may hope that also the gauged BLG theory  exists and that the full theory can be found\footnote{ The complete theory has now been constructed, see  \cite{Gran:2012}.}.

One can also in this case consider turning off the four-indexed structure constants and reduce the theory to that relevant for one brane. This theory  should then, based on the degrees of freedom argument above, contain a potential with  a similar 
structure as in the ABJM case. At least we  expect  the potential in the topologically gauged BLG theory to contain 
 "triple trace"  terms 
 which give rise  to a Higgs effect leading to a chiral point TMG theory and  properties (physical modes etc) similar to those  in the topologically gauged ABJM case.

   \subsection{Scaling limits and gravity free matter theories in curved backgrounds}
   
   Above we have discussed the result of the higgsing and checked explicitly that it leads to a chiral point supergravity of the TMG type. 
   This statement was seen to be true
   for any value of the scalar VEV $v$, level $\lambda=\tfrac{2\pi}{k}$ and $g_M$ introduced in the previous section. Thus it is of some interest to form physical constants by considering various 
   combinations of these three parameters and to see how the physics depends on the scaling of the scalar VEV to infinity which in the ungauged case corresponds to going from M-theory
   to D2-branes in string theory. In fact, the higgsing done here turns out to be a rather straightforward generalization of the Higgs effect for Chern-Simons theories in three flat dimensions originally found by Mukhi and Papageorgakis \cite{Mukhi:2008ux}. In  \cite{Chu:2010fk} we follow basically  the same steps  for our higgsed topologically gauged M2-theories with six supersymmetries and 
   arrive at a lagrangian with  the following structure, before taking the scaling limit,
   \beqa
   L&&=\tfrac{1}{\mu\kappa^2}L_{CD(\om)}-(\tfrac{1}{\kappa^2}+\tfrac{z}{\kappa}+z^2) R+\tfrac{1}{g_{YM}^2} (F^+)^2-DzD\bar z\cr
   &&-(g_{YM}^2 z^4+subleading)-(\mu g^{}_{YM}z^3+subleading)-(\tfrac{\mu^2}{\kappa^2}+subleading),
   \eeqa
   where we have used the definitions
   \beq
   \kappa^2\propto \tfrac{1}{v^2},\,\,\,g^2_{YM}\propto (\lambda v)^2,\,\,\,\mu\propto(g_Mv)^2.
   \eeq
   
   We now need a strategy to get a sensible result in the scaling limit $\lambda\rightarrow 0$ corresponding to the compactification circle shrinking to zero radius. Demanding that $g_{YM}$ stays fixed  means that $v\rightarrow \infty$.  If we also want  a fixed cosmological constant,
   which we saw above behaves as $\Lambda\propto \mu^2 \propto (g_Mv)^4$, we need also  $g_M\rightarrow 0$. This determines the scaling behaviour of  all three parameters
   in the original lagrangian: $\lambda\propto v^{-1}$, or in fact the level goes to infinity as $k\propto v$, while $\kappa$ and $ g_M$  go to zero as $\propto v^{-1}$. In the quiver version of this analysis carried out in \cite{Chu:2010fk} also $N$ (from $U(N)\times U(N)$) enters and the 't Hooft parameter can be introduced.
   
   This means that as the VEV diverges the AdS geometry stays fixed while the gravitational coupling constant $\kappa$ goes to zero. Note that the unwanted linear term in the fluctuations of the scalar fields $z$ that appears in the above lagrangian actually cancels a term that is subleading to the cosmological term on the last line. The end-result is thus a non-conformal matter theory
   consisting of scalars, spinors and Yang-Mills fields with six supersymmetries living in a fixed $AdS$ geometry without gravity. Other methods to construct matter theories living in fixed non-trivial (i.e. not flat) geometries have been discussed recently
   by Festuccia and Seiberg  \cite{Festuccia:2011ws}, see also \cite{Jia:2011hw}. 
   
   It might be interesting to note \cite{Chu:2010fk} that if the procedure just described had been carried out starting with the above  lagrangian but now multiplied by $g_M^2$
   the scaling limits inferred from taking the VEV $v$ to infinity would be rather different. In this case both $g_{YM}$ and Newton's constant $\kappa$ can be kept fixed
   while the geometry becomes infinitely curved if we try to decouple gravity. The final result is thus in this case a  supergravity theory that seems to  make sense only
   for non-zero values of Newton's constant.
   
   \subsection{Indications of a "sequential AdS/CFT" phenomenon}

%\subsection{Topologically gauged M2-branes:  a first step towards\\ "sequential AdS/CFT"?}

The discussion in this section is based on the ABJM results  presented in the previous section but relies heavily also on some more speculative points to be made precise below.
We will try to argue that an explicit realization of what we will call "sequential AdS/CFT" might follow from the higgsing properties of topologically gauged ABJM theories
explained above\footnote{This should also apply  to the topologically gauged BLG theory that has now been constructed \cite{Gran:2012}.}.  This will require at least the following three assumptions: (1) topologically gauged M2 brane theories arise in $AdS_4/CFT_3$ from
adopting Neumann boundary conditions instead of the usual Dirichlet ones, (2) it is possible to make sense of "going to the boundary 
twice"\footnote{Techniques like those discussed in \cite{Sati:2012tt} may be useful in this context.}, perhaps by relating the higgsing 
(and the breaking of the  symmetries of the $CFT_3$   to those  of $AdS_3$)  to a change of foliation of $AdS_4$ and (3)
the TMG chiral supergravity  in $AdS_3$ can be used in an $AdS_3/CFT_2$ context and thus has some $CFT_2$ on the boundary.

Some arguments in favor of these assumptions can be found in the literature although conclusive ones may not exist for any of the three assumptions.  
We will here discuss the assumptions  in the  context of topologically gauged ABJM theories in order to
assess the possibilities to realize the "sequential AdS/CFT" in the specific case
\beq
AdS_4(N)/CFT_3(TG)\rightarrow AdS_3(H)/CFT_2
\eeq
where $N$ refers to Neumann boundary conditions, $TG$ to a topologically gauged ABJM/ABJ (or BLG) theory and $H$ to its higgsed version. The final $CFT_2$ is unknown.
Needless to say, if this sequence  can be made sense of, a 
further extension to higher or lower dimensions would be very interesting (see e.g. \cite{Strominger:1998yg, Sen:2011cn,Dabholkar:2011ec}). Ideas somewhat related to those discussed here can be found in e.g. 
\cite{Compere:2008us} and more recently in \cite{Andrade:2011nh}. We will have reason to come back to the last work again below. There is also a speculation due to M. Vasiliev \cite{Vasiliev:2001zy} based on algebraic higher spin arguments about
the possibility to have more than one $AdS/CFT$ following each other. However, all  previous  proposals   lack a dynamical realization or mechanism for connecting
two or more $AdS/CFT$ correspondencies. The new ingredient  here that perhaps can make this scenario more realistic is the conformal spontaneous symmetry breaking, ending in a theory in $AdS_3$, triggered by the new terms in the scalar field potential
generated by the topological gauging.  Note that this discussion is relevant for stacks of branes as well as for a single one.

We now discuss the three assumptions in succession:\\
\\
{\it Assumption 1: Topologically gauged M2 brane theories arise in $AdS_4/CFT_3$ from
adopting Neumann boundary conditions instead of the usual Dirichlet ones.}\\
\\
The issue of which boundary conditions to use and their relation to the terms that appear in the expansion of bulk fields in powers of the radial coordinate
has been discussed both for scalars within the Breitenlohner-Freedman bound \cite{Witten:2001ua} and outside \cite{Marolf:2006nd}, as well as for Chern-Simons vector fields in three dimensions  \cite{Witten:2003ya}. Unitarity is a central and unsolved issue in most  discussions on Neumann boundary conditions, see for instance
 \cite{Andrade:2011dg} and references therein. In the context of $AdS_3/CFT_2$ that will be addressed below, see also \cite{Andrade:2011sx}.  Attempts to generalize these results for spin 0 and 1 to spin 2 can be found in several papers where for instance
the relation of the Cotton tensor to Neumann boundary conditions in $AdS_4/CFT_3$ is pointed out. Among the  more explicit papers are  \cite{Compere:2008us, deHaro:2008gp, Amsel:2009rr}\footnote{A possible connection to topologically gauged BLG theory was in fact mentioned already in the last of these references.}. In fact, one may introduce  \cite{deHaro:2008gp} two  CFT's of opposite chirality as boundary field theories dual to the same $AdS_4$ bulk theory with Neumann boundary conditions which gives a possible explanation of the arbitrariness in the choice of chirality of the conformal gravity sector in the topological gauging procedure.

Concerning the relation between the Dirichlet and Neumann in the spin two case, one needs to understand how the interpretation of the
zeroth and third order terms in the radial expansion of the Fefferman-Graham  metric can be interchanged. Thus consider the metric (see e.g. \cite{deHaro:2008gp})
\beq
ds^2=\frac{l^2}{r^2}(dr^2+(\eta_{ij}+h_{ij}(r,x))dx^idx^j),
\eeq
which, when inserted into the $AdS_4$ gravitational bulk equations,  leads to an expansion in the radial coordinate
\beq
\bar h_{ij}(r,p)=(1+...)h^{(0)}_{ij}(p)+((pr)^3+...)h^{(3)}_{ij}(p).
\eeq
As explained in \cite{deHaro:2008gp} using boundary conditions
\beq
{\Box}^{1/2}h^{(0)}_{ij}=\pm \epsilon_{ikl}\partial_k h^{(0)}_{lj}
\eeq
 makes it possible to define a Legendre transformation  related to the Cotton tensor. This Legendre transformation is supposed  to take one from  a (free) $CFT_3$ in the UV related to Dirichlet boundary conditions to two new (interacting) $CFT_3$'s 
 of opposite chirality in the IR related to Neumann boundary conditions. What is needed for this to work is a third order derivative  operator that can be inverted in the same sense as the 
 Chern-Simons operator appearing in the usual superconformal M2 theories or in the work of Witten in \cite{Witten:2003ya}. 
 That  the linearized Cotton tensor has the wanted  properties  is 
demonstrated in \cite{deHaro:2008gp}.
This may also be seen by solving for all the components of the metric from the linearized inhomogeneous Cotton equation written out in the light-cone gauge \cite{Nilsson:2008ri}: 
\beqa
h_{++}&=&-2\partial^{-3}_-T_{2-},\cr
h_{+2}&=&-\partial^{-3}_-T_{--},\cr
 \partial_+h_{++}&=&2\partial^{-2}_-T_{2+}-2\partial^{-3}_-(\partial_2T_{22}),\cr
 \partial_+h_{+2}&=&\frac{1}{2}\partial^{-2}_-T_{22}-\partial^{-3}_-(\partial_2T_{2-}),\cr
 \partial_+^2h_{+2}&=&-\partial^{-1}_-T_{++}+\partial^{-3}_-(\partial_2^2T_{22}),
\eeqa
where $T_{\mu\nu}$ with $\mu=+,-,2$ is the light-cone components of the stress tensor for the matter system in question.
This, in fact, also shows that the topologically gauged theories do not contain any new propagating degrees of freedom in three dimensions as a result of the gauging.  The spin two Legendre transformation
\beq
\tilde W(\tilde h)=W(h)+V(\tilde h,h)
\eeq
is generated by the second variation of the gravitational Chern-Simons term
\beq
V(\tilde h,h)=-\tfrac{l^2}{2\kappa^2}\int d^3x h_{\mu\nu}\frac{\de^2 S_{(CS)}}{\de g_{\mu\nu}\de g_{\rho\si}}\tilde h_{\rho\si},
\eeq
where $h$ is the metric flucuation in 
the Dirichlet case  and $\tilde h$ in the Neumann case. $W$ and $\tilde W$ are the on-shell actions in the two cases.

The situation described above should be compared to the simpler case of the $O(N)$ model where things are better controlled \cite{Sundborg:2000wp, WittenatJHS60:2001, Sezgin:2002rt, Klebanov:2002ja, Koch:2010cy,Jevicki:2011ss}. The UV fixed point
is then described by a free theory while the non-trivial fix point is in the IR and interacting but weakly so. Furthermore,  there is an explicitly  known Legendre transformation \cite{Girardello:2002pp, Petkou:2003zz} between these two fix points theories whose bulk dual is also understood to some extent in terms of Vasiliev's  higher spin (HS) theory \cite{Vasiliev:1999ba}.
 It is 
interesting to note that the HS bulk theory dual to the free UV $CFT_3$ has massless gauge fields of all spins from two and up while the interacting IR $CFT_3$ is a higgsed HS theory where all fields with spin above two have become massive \cite{Girardello:2002pp,Giombi:2009wh, Giombi:2011ya}. Recent work supporting this picture can be found, e.g., in \cite{Maldacena:2011jn}.

More recently other indications of HS duals of interacting fix point theories have appeared in \cite{Aharony:2011jz}. Here one is analyzing
scalar $O(N)$ and $U(N)$ models in the singlet sector, implemented by coupling the theory to a gauge theory of Chern-Simons type in order to add only trivial 
operators\footnote{This is the same argument  as used  in this paper for gravitational theories.}.
In perturbation theory in the 't Hooft coupling $\lambda=4\pi\frac{N}{k}$ one finds that all the HS currents are preserved at infinite $N$ even at the interacting fix point although they
are broken by $\frac{1}{N}$ effects. Related results are found for fermions in \cite{Giombi:2011kc}. The interesting implication of these results seems to be that
there should exist new HS theories beyond the known Vasiliev type theories in $AdS_4$  \cite{Vasiliev:1999ba,Sezgin:2002ru} related to the known ones by some kind of deformation.  Another intriguing result in this context is the appearance of lines of fixed points similar to what generically happens in two-dimensional conformal field theory. For a recent account of some
 higher spin issues in this context, see \cite{Bekaert:2012ux}.

In the case of the $O(N)$ models with their weak-weak duality one may actually hope to be able to prove the $AdS/CFT$ correspondence. One such attempt
is advocated in \cite{Koch:2010cy, Jevicki:2011ss} based on bilocal "collective fields" following on ideas that started with Sundborg \cite{Sundborg:2000wp}, Witten \cite{WittenatJHS60:2001},  and Sezgin and Sundell 
\cite{Sezgin:2002rt}. \\
\\
{\it {
Assumption 2: The symmetry breaking  occurring in the topologically gauged  ABJM/ABJ theories  bringing the theory from a $CFT_3$ 
to a (chiral) TMG theory in $AdS_3$ corresponds to a change of foliation in $AdS_4$ 
 }}
\\
\\
The higgsing that takes place in the topologically gauged M2-brane theories described in the previous section breaks the three-dimensional superconformal symmetry to that of $AdS_3$. This phenomenon\footnote{This symmetry breaking can also be viewed as the result of a gauge choice as shown in \cite{Marnelius:1979pr}.} was noticed in \cite{Chu:2009gi}, and analyzed in detail for some specific cases in \cite{Chu:2010fk}. It follows from combining the effect of the conformal coupling term of two scalar fields and the scalar curvature with the last set of potential terms that are independent of the structure constants.  What is found is that  the theory has an $AdS_3$ solution that sits exactly at  a chiral point
similar to that discussed by Li, Song and Strominger \cite{Li:2008dq} but with signs corresponding to TMG. This changes the sign of the  energies of the black holes and the would-be propagating physical gravity modes away from the chiral point. This thus  leads to the appearance of negative energy black holes which  constitute a potential problem for the theories discussed here. What this means for the log-gravity/cft issue if anything is not clear.

The higging also involves finding the correct super-$AdS_3$ symmetry generators after breaking as combinations of the ones in the superconformal algebra. By checking the supersymmetry transformation rules after breaking one can infer the answer. One finds for instance that the covariant derivative in the conformal theory gets augmented by a gamma matrix term in a way that is familiar from any supergravity theory in AdS. This way of relating conformal and AdS symmetry algebras has been discussed, e.g., by M. Vasiliev in
\cite{Vasiliev:2001zy} based on HS algebra arguments, however, without having a theory exhibiting an appropriate Higgs phenomenon like the one we find in the topologically gauged ABJM theories constructed in  \cite{Chu:2009gi}.

In the context of $AdS_4/CFT_3$ the symmetry breaking taking place in the boundary theory requires an interpretation in the $AdS_4$ bulk theory. One possible such interpretation could be that it corresponds to a shift of foliation in the bulk. In fact it is
well known that one can foliate $AdS$ spaces in a number of different ways some leading directly to a boundary also with $AdS$ geometry \cite{Emparan:1999pm}. For an explicit application to $AdS_4$, see \cite{Jatkar:2011ue} where for instance the following   case is discussed:
\beq
ds^2=\frac{1}{k+\frac{r^2}{L^2}}dr^2+(k+\frac{r^2}{L^2})L^2d\Omega^2+r^2d\tilde\Omega^2,
\eeq
where  $k=\pm 1$ and  $d\Omega^2$ and  $d\tilde\Omega^2$  are  metrics that can be either on a  spherical or a hyperbolic space of arbitrary dimensions. This metric  may be compared to the usual $AdS/CFT$ metric which if written 
with the same radial coordinate reads
\beq
ds^2=\frac{L^2}{r^2}dr^2+r^2dx_i^2.
\eeq
Here the index on the coordinates $x_i$ extends over the number of flat dimensions corresponding to both the  spherical and hyperbolic spaces above. There are a number of questions concerning the nature of the boundaries in these cases and how they relate to each other. In fact, non-standard foliations could be problematic for unitarity, see 
\cite{Nakayama:2012sn}.\\
\\
{\it Assumption 3: The TMG chiral supergravity  makes sense in the  $AdS_3/CFT_2$ context and thus has a $CFT_2$ on the boundary}\\
\\
The kind of $CFT_2$ that would be relevant for topologically gauged M2-branes is probably not yet discussed in the literature. We will therefore
just mention some results that are well-known and that hopefully will have  generalizations useful in our context. The question of the role of the chiral point in $AdS_3/CFT_2$ has been analyzed in the 
non-supersymmetric case by Skenderis et al in \cite{Skenderis:2009kd} where  also the issue of
logarithmic modes was addressed. In a more general context there has  recently been some important progress  \cite{Bergshoeff:1989ns, Gaberdiel:2010pz, Chang:2011mz,  Chen:2011yx} concerning an $AdS/CFT$ connection between ${\mathcal W}_N$ $CFT$'s in two dimensions and HS theories in $AdS_3$  of the Vasiliev type, see e.g. 
\cite{Prokushkin:1998bq, Shaynkman:2001ip}. The massless HS fields play here a crucial role since they correspond to the set of conserved HS currents
in the ${\mathcal W}_N$ CFT's, where $N$ can be any finite number $\geq 2$ in this case. These Vasiliev type HS theories contain also two free massive scalar fields which make the theory non-conformal. It might be interesting to investigate if there is also here a conformal phase in the bulk and what field theory it might correspond to on the boundary.

Having discussed if the higgsed topologically gauged ABJM theory in $AdS_3$ may have a conformal boundary field theory in two dimensions, one could also ask
if  this theory in three dimensions can be associated with a boundary field theory even before higgsing, that is, as a conformal theory in three dimensions.
Questions of this kind have indeed been addressed recently in the simpler setting of pure conformal gravity, see   \cite{Afshar:2011yh, Afshar:2011qw}.

\acknowledgments

I would like to thank X. Chu, H. Nastase, J. Palmkvist, C. Papageorgakis, P. Sundell and M. Vasiliev for stimulating discussions. In particular, I am grateful  to
Per Sundell for reading parts of the manuscript and sharing his insights in higher spin theory with me. This note
contains results presented at recent conferences in Budapest, Prague, Nordita and Bilbao. I would also like to thank the people at
Center for the Fundamental Laws of Nature at Harvard for their kind hospitality
during my short visit there in September 2011 in connection with a seminar where parts of this material was presented.
I also thank the Theoretical Physics Group at the University of New Hampshire for a very nice visit. The work is partly funded by the Swedish
Research Council.

\end{document}